\documentclass{llncs}
\usepackage{mathptmx}       
\usepackage{helvet}         
\usepackage{courier}        
\usepackage{type1cm}        
\usepackage{makeidx}         
\usepackage{graphicx}        
\usepackage{multicol}        
\usepackage[bottom]{footmisc}
\usepackage{subfigure}
\usepackage{amsfonts}
\usepackage{adjustbox}
\usepackage[cmex10]{amsmath}

\begin{document}

\title{Spectral Multi-scale Community Detection in Temporal Networks with an Application}
\titlerunning{TMSCD Application}  
\author{Zhana Kuncheva \inst{1,2} \and Giovanni Montana \inst{2,3}}
\authorrunning{Kuncheva et al.} 
\institute{
Clinical Development and Mathematics, C$4$X Discovery Ltd, M1 $3$LD, UK.\and
Department of Mathematics, Imperial College London, SW$7$ $2$AZ, UK.\\
\email{z.kuncheva12@imperial.ac.uk}\and
Department of Biomedical Engineering, King's College London, SE$1$ $7$EH, UK.\\
\email{giovanni.montana@kcl.ac.uk}
}
\maketitle              
\section{Introduction}
The analysis of temporal networks~\cite{Holme2012} has a wide area of applications in a world of technological advances. An important aspect of temporal network analysis is the discovery of community structures~\cite{Newman2006}. Real data networks are often very large and the communities are observed to have a hierarchical structure referred to as multi-scale communities. Changes in the community structure over time might take place either at one scale or across all scales of the community structure. 

The multilayer formulation~\cite{Kivel} of the \textit{modularity maximization} (MM) method~\cite{Newman2006} introduced in \cite{Mucha} captures the changing multi-scale community structure of temporal networks. This method introduces a coupling between communities in neighboring time layers by allowing inter-layer connections, while different values of the resolution parameter $\gamma$ enable the detection of multi-scale communities. However, the range of parameter values $\gamma$ must be manually selected. When dealing with real life data, communities at one or more scales can go undiscovered if appropriate parameter ranges are not selected. 

Our recent work on multi-scale community detection in temporal networks~\cite{Kuncheva2017c} proposes a novel Temporal Multi-scale Community Detection (TMSCD) method, which overcomes the obstacles mentioned above. This is achieved by using the spectral properties of the temporal network represented as a multilayer network. In this framework we select automatically the range of relevant scales within which multi-scale community partitions are sought. 
\section{The Method of Temporal Multi-scale Community Detection (TMSCD)}
The proposed TMSCD method relies upon the notion of spectral graph wavelets~\cite{Hammond2009}, and is a multilayer
extension of the multi-scale community detection procedure via spectral graph wavelets developed in \cite{Tremblay2014}.

First, we build a multilayer representation of the temporal network considering new inter-layer weights connecting nodes in neighboring time layers. This introduces dependence between neighboring time points. Second, we apply the definition of a spectral graph wavelet~\cite{Hammond2009} at every node for each time layer. The main idea of the spectral graph wavelet approach is that wavelets at small scales span the local neighborhood of nodes, while wavelets at larger scales span an increasing number of neighboring time layers. 

The most essential part in~\cite{Hammond2009} is the design of a wavelet filter function $g$. The crux of the TMSCD method is the construction of a $B$-spline based wavelet filter function $g$ adapted for multi-scale community detection
in temporal networks. By arguments from Perturbation theory~\cite{Bhatia1997} we consider the separate layers as disconnected components, while the inter-layer weights as perturbations. In this way, when studying
the spectral properties of the supra-Laplacian (the Laplacian of the multilayer formulation), we take into account the fundamental difference between within-layer and inter-layer edges.

By Perturbation theory the eigenvectors corresponding to the smallest eigenvalues of the supra-Laplacian are linear combinations of the eigenvectors 
(corresponding to the $0$ eigenvalues) of the Laplacian matrices of the separate time layers. From spectral graph theory~\cite{Chung1996}, it is known that an eigenvector corresponding to the $0$ eigenvalue of the Laplacian matrix is not informative of the community structure. For this reason, the eigenvectors of the supra-Laplacian matrix, which can be obtained as approximations of these eigenvectors, cannot be used to identify within-layer communities. Hence, we propose a procedure for the selection of certain larger eigenvalues, which are ``informative'' of the prevalent community structure over time. Thus we reconsider the role of the Fiedler vector in community detection for temporal networks.     

The contribution of the eigenvectors at different scales $s$ is controlled by a $B$-spline based wavelet filter function $g$, whose few parameters are automatically selected. The filter puts more weight to smaller ``informative'' eigenvalues when large scale communities are sought, and more weight to larger eigenvalues when small scale communities are sought. 

A series of simulations on various benchmarks presented in~\cite{Kuncheva2017c} show the competitive performance of TMSCD to MM. These simulations show that the proposed inter-layer weights perform better than fixed inter-layer weights. Another advantage of TMSCD over MM is the automated selection of scales' ranges at which multi-scale communities should be sought. 
\section{Application to Primary School Data}
Here we present for the first time initial results from applying the TMSCD method to the temporal social patterns appearing in a primary school~\cite{Stehle2011a}.
Data on face-to-face interactions between $242$ students and teachers from $10$ classes were collected for two days, which we subdivide into $36$ intervals of $30$ minutes. For each interval, the network of interactions has a link between two individuals if those individuals had at least one contact during the corresponding interval. 

\begin{figure}[htb]
\centering

\subfigure[Instability of detected communities at each scale $s$ with significance threshold.]{\includegraphics[width=.48\columnwidth,height=0.25\textheight]{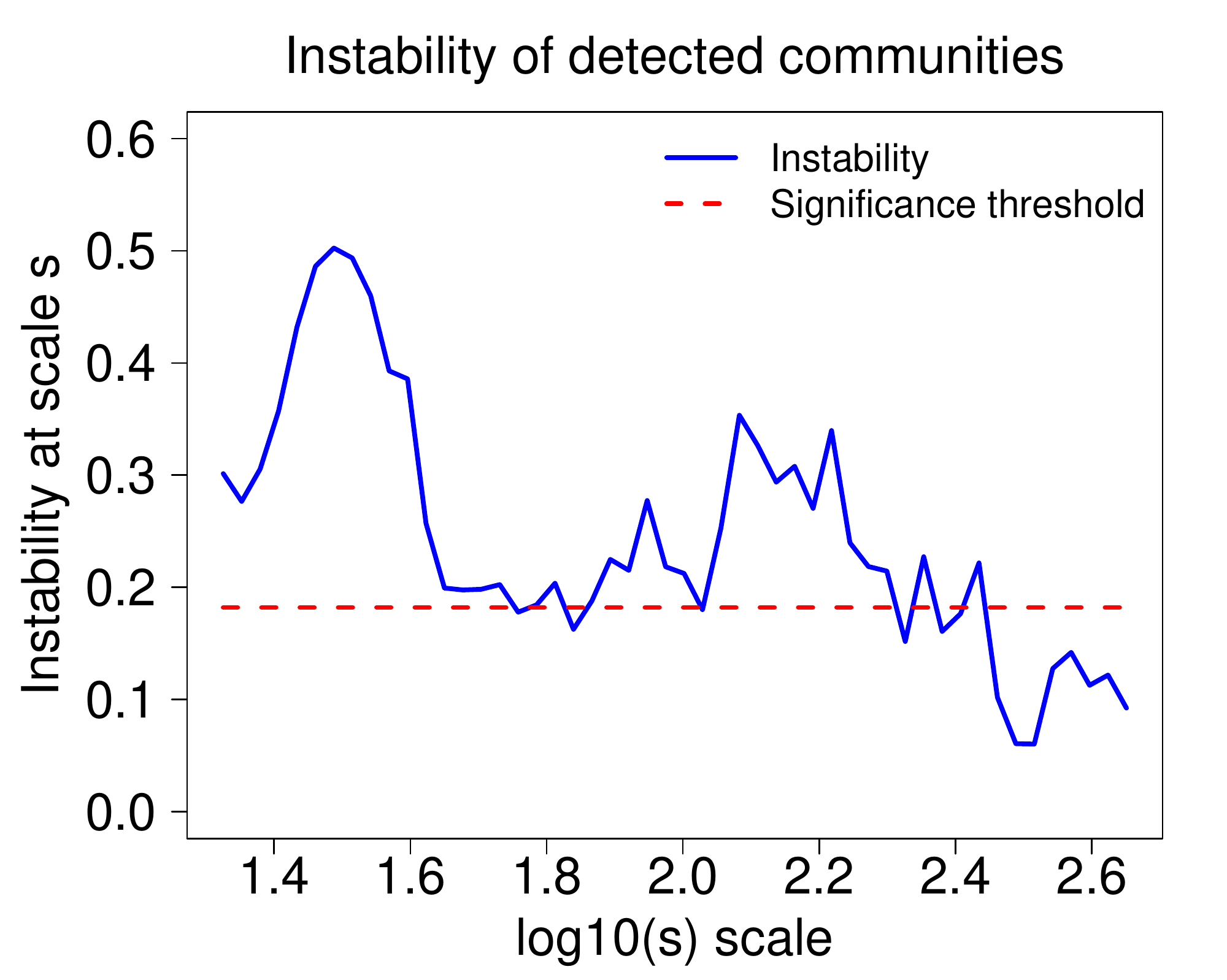}\label{fig:numbPrimary:a}}
\hfill
\subfigure[Number of communities and singletons at each time point for two different scales.]{\includegraphics[width=.51\columnwidth,height=0.25\textheight]{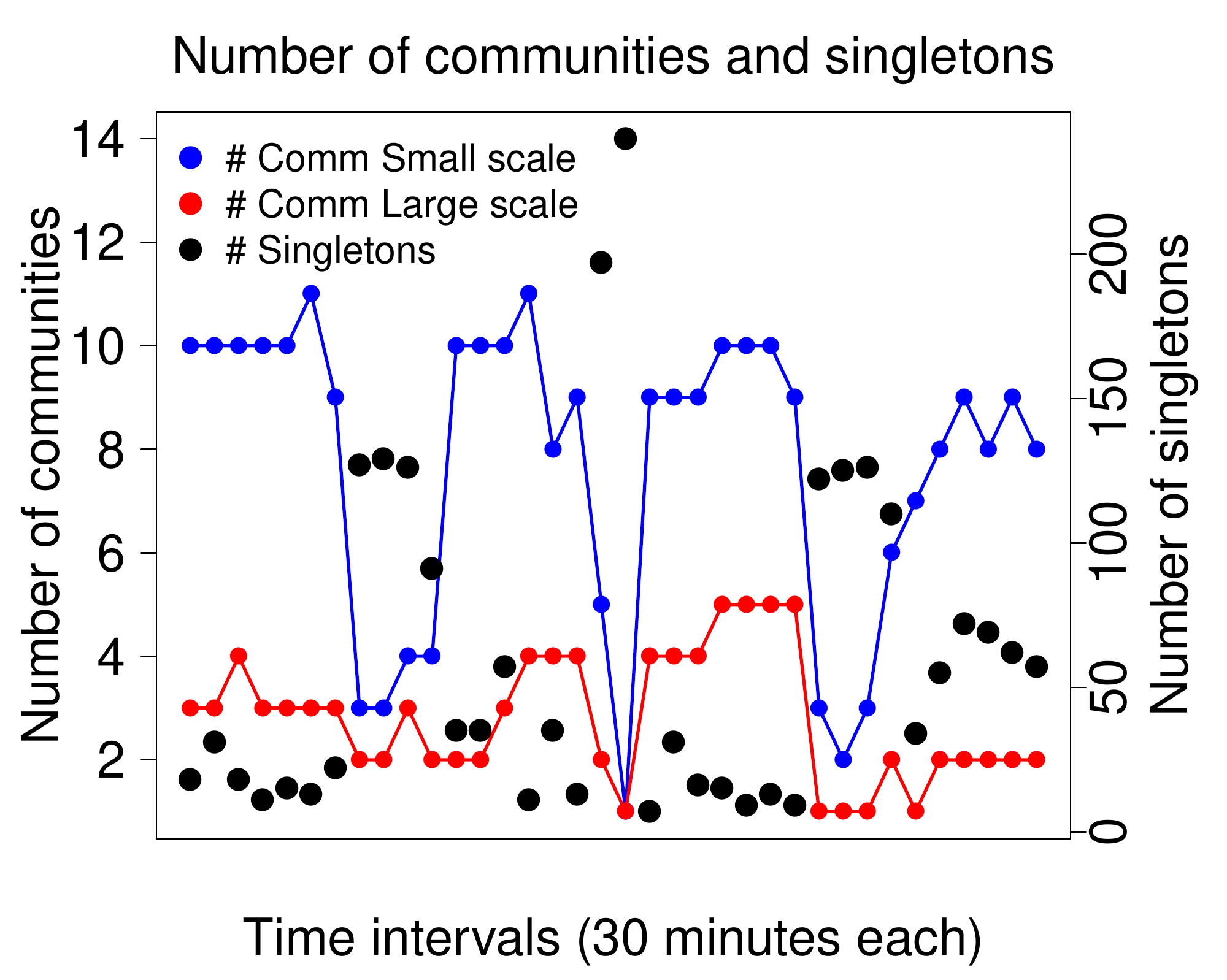}\label{fig:numbPrimary:b}}
\caption{Results from applying TMSCD to temporal social patterns appearing in a primary school.}
\label{fig:numbPrimary}
\end{figure}

After applying the TMSCD method to this temporal network, we assess the statistical significance of the communities at each scale, which identified two stable scales of partitions, Fig.~\ref{fig:numbPrimary:a}. For the number of communities at each time point for the two visualized partitions and the number of singletons see Fig.~\ref{fig:numbPrimary:b}.

For the smaller scale, consecutive time intervals which overlap with pre- and after-lunch class periods group students in around $9-10$ communities. This means that each class is in its own community and communication takes place between class peers. Fewer defined communities appear during lunch and at the beginning/end of the school day when many students walk around and no long face-to-face contact is observed. The large scale captures the split between lower grades (first, second and third) and higher grades (fourth, fifth and sixth), or between lower (first and second), middle (third and fourth) and higher (fifth and sixth) grades. Overall, these results demonstrate the strength of TMSCD to detect multi-scale communities in temporal networks. Future work involves more thorough real life applications and improvements of the scalability issues of the method. The author ZK acknowledges partial support by grant no. I $02/19$ of Bulgarian NSF.
{\scriptsize
\bibliographystyle{splncs03}
\bibliography{Lib}
}
\end{document}